\documentclass{INTERSPEECH2023}
\usepackage{multirow}
\newcommand{\specialcell}[2][l]{%
  \begin{tabular}[#1]{@{}l@{}}#2\end{tabular}}
\usepackage[acronym]{glossaries}
\usepackage{booktabs}
\newacronym{MAE}{MAE}{mean absolute error}
\newacronym{GT}{ground truth}{ground truth}
\usepackage{amsmath,amssymb}
\DeclareMathOperator{\E}{\mathbb{E}}
\usepackage[subtle]{savetrees}
\interspeechcameraready

\usepackage{todonotes}

\title{Emo-StarGAN: A Semi-Supervised Any-to-Many Non-Parallel Emotion-Preserving Voice Conversion}
\name{Suhita Ghosh$^{1*}$, Arnab Das$^{1,3*}$, Yamini Sinha$^2$, Ingo Siegert$^2$, Tim Polzehl$^3$, Sebastian Stober$^1$}
\address{
  $^1$Artificial Intelligence Lab (AILab), Otto-von-Guericke-University, Magdeburg, Germany\\
  $^2$Mobile Dialog Systems, Otto-von-Guericke-University, Magdeburg, Germany \\
  $^3$Speech and Language Technology, German Research Center for Artificial Intelligence (DFKI)
  }
\email{\{suhita.ghosh,yamini.sinha,ingo.siegert,stober\}@ovgu.de \\
\{arnab.das,tim.polzehl\}@dfki.de}

\begin{document}

\maketitle
 \def\thefootnote{*}\footnotetext{These authors contributed equally to this work}\def\thefootnote{\arabic{footnote}}
\begin{abstract}
Speech anonymisation prevents misuse of spoken data by removing any personal identifier while preserving at least linguistic content.
However, emotion preservation is crucial for natural human-computer interaction.
The well-known voice conversion technique StarGANv2-VC achieves anonymisation but fails to preserve emotion.
This work presents an any-to-many semi-supervised StarGANv2-VC variant trained on partially emotion-labelled non-parallel data.
We propose emotion-aware losses computed on the emotion embeddings and acoustic features correlated to emotion.
Additionally, we use an emotion classifier to provide direct emotion supervision.
Objective and subjective evaluations show that the proposed approach significantly improves emotion preservation over the vanilla StarGANv2-VC.
This considerable improvement is seen over diverse datasets, emotions, target speakers, and inter-group conversions without compromising intelligibility and anonymisation.

\end{abstract}

\section{Introduction}
The increasing use of cloud-based speech devices, such as smart speakers, raises concerns about the protection and confidentiality of the sensitive data being collected and used~\cite{wienrich2021trustworthiness,HaaseHCII2022}.
In case of data compromise, the spoken data can be exploited to bypass the speaker verification systems or impersonate authorised users~\cite{RUB35,RUB66}.
This makes it crucial to anonymise the utterance before being shared across systems, such that the speaker cannot be traced back.
Voice conversion (VC) achieves anonymisation by modifying the utterance of the source speaker to sound like another target speaker while preserving at least linguistic content.
In cases where the response of a speech device is driven by the end-user's emotional state, the preservation of emotion also becomes pertinent, e.g., a digital assistant responding with comforting words when the user sounds sad.

Many VC approaches using parallel data have been proposed, such as parametric statistical modelling-based~\cite{Kobayashi2016TheNV,helander2011voice}, non-parametric exemplar-based~\cite{takashima2013exemplar,jin2016cute} and deep neural network-based~\cite{liu2020voice}.
Parallel data comprise utterances having the same linguistic content from both the source and target speakers, which is arduous and expensive to acquire.
Therefore, recent works focus more on non-parallel data, as it is simpler to obtain and better represents real-life situations where any arbitrary speech requires anonymisation.

A few non-parallel VC approaches~\cite{lian2020arvc,sun2016phonetic} use
phonetic posteriorgrams (PPGs) as one of the inputs to the encoder-decoder framework to generate translated acoustic features.
These methods tend to produce mispronunciations due to alignment issues~\cite{liu2021any}, resulting in degraded prosody, which provides cues about emotion~\cite{cao2014prosodic}.
The non-parallel variational autoencoder (VAE) approaches~\cite{wu2020vqvc+,tang2022avqvc} typically disentangle the content and speaker embeddings using a reconstruction loss and relevant constraints to remove speaker information.
The VAE-based approaches are prone to spectrum smoothing, which leads to a buzzy-sounding voice, dampening the emotion~\cite{sisman2020overview}.
A plethora of generative adversarial network (GAN) based VC approaches~\cite{sisman2020overview,sakamoto2021stargan} were proposed, which can use non-parallel data due to cycle-consistency loss~\cite{kaneko2018cyclegan}.
GANs overcome the over-smoothing effect through a discriminator, which teaches the generator to produce natural sounding conversions.
Recently, StarGANv2-VC~\cite{li2021starganv2vc}, a non-parallel any-to-many GAN-based VC technique has been proposed.
The method is attractive due to its fast real-time conversion and naturally sounding samples with high intelligibility.
However, the model fails to preserve the source speaker’s emotion, especially for diverse emotions and acoustic conditions such as high varying pitch. \\
\indent Thus, we propose the novel ``Emo-StarGAN'' in this paper, which is an any-to-many semi-supervised \textit{emotion-preserving} variant of StarGANv2-VC.
Two kinds of emotion supervision are proposed: (i) \textit{direct}: through an emotion classifier, which provides feedback to the generator when the emotion ground truth is available. (ii) \textit{indirect}: through losses computed between source and conversions using emotion embeddings or acoustic descriptors correlated with emotion, improving the conversion quality for diverse target speakers.
Extensive evaluation is conducted on three datasets, diverse target speakers, emotions, and various group conversions such as accent and gender.
Both objective and subjective evaluations portray that Emo-StarGAN improves emotion preservation significantly over StarGANv2-VC for all cases, without hurting the naturalness, intelligibility and anonymisation.
\begin{figure}[!t]
  \centering
  \includegraphics[scale=1.05]
  {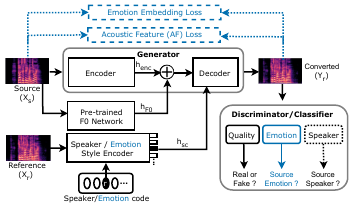}
\vspace{-1em}
  \caption{The proposed framework adapted from StarGANv2-VC~\cite{li2021starganv2vc}. The blue components do not belong to StarGANv2-VC. In voice conversion, the style encoder captures speaker embeddings.
  The same framework is used for emotion conversion, where the style encoder learns emotion embeddings.
  The dashed components are not used in the emotion embedding training.}
  \label{fig:architecture}
\vspace{-1em}
\end{figure}
\section{StarGANv2-VC Architecture}
\label{sec:sgvc}
Our method is based on the StarGANv2-VC architecture, as shown in Figure~\ref{fig:architecture}.
A \textit{single} generator $G$ is trained to convert a source utterance $X_s$ to the target utterance $Y_r$, conditioned on the speaker style embedding $h_{sc}$.
The speaker style embedding $h_{sc}$ represents \textit{speaker characteristics}, such as accent.
The speaker style-encoder $SE$ produces the speaker style embedding $h_{sc}$ using the target speaker's mel-spectrogram $X_r$ having the style information and, target speaker's code $r$ (one-hot encoding).
$SE$ comprises multiple convolutional layers which are shared by all the speakers, followed by a speaker-specific linear projection layer, which outputs an embedding $h_{sc}$ for each target speaker.
A mapping network~$M$ having the same architecture as $SE$  is trained along with it, which inputs a random latent vector instead of a reference mel-spectrogram, providing diverse style representation for all speakers.
The converted sample produced by the generator $Y_r$ = $G(X_s, h_{F0}, h_{sc})$ captures the style of the target speaker-code $r$ and has the linguistic content of the source utterance $X_s$.
In order to produce F0-consistent conversions, the generator is fed with source-pitch embedding $h_{F0}$ along with source utterance $X_s$ and style representation $h_{sc}$.
The pitch embedding $h_{F0}$ is derived from the convolutional outputs of a pre-trained F0 network~\cite{kum2019joint}.
The framework consists of one discriminator $D$ and one adversarial source speaker classifier $C_s$.
$D$ is the typical adversarial discriminator, which encourages the generator to produce plausible conversions.
$C_s$ has the same architecture as $D$, which is trained to enforce the generator to produce conversions having no details about the source speaker.

\section{Emo-StarGAN}
\label{sec:ours}
Recent VC works~\cite{walczyna2023overview} including StarGANv2-VC have primarily focused on generating naturally sounding voices with correct linguistic content, and not much on emotion preservation.
The proposed Emo-StarGAN aims to anonymise an utterance by modifying the source speaker's timbre, while preserving the source's linguistic and \textit{emotional} content $e_s$.

\subsection{Direct Emotion Supervision}
Our framework uses an additional emotion classifier $C_{e}$ which provides \textit{direct} emotion supervision for utterances having emotion labels, as shown in Figure~\ref{fig:architecture}.
$C_{e}$ encourages the generator to produce \textit{emotion-consistent} samples, such that the source and target samples have the same emotion.
When $C_{e}$ is trained, the generator weights are fixed, and the emotion classifier is trained to ascertain the emotion of the source utterance through the classification loss $L_{emod}$.
\begin{equation}
\vspace{-0.2em}
\label{eqn:emod}
 L_{emod} = \E_{X_s, e_s}\big[CrossEntropy(C_{e}(X_s), e_s)\big]
 \vspace{-0.2em}
\end{equation}
In contrast, during the training of the generator, $C_{e}$ weights are fixed, and the generator is encouraged to produce samples having the same emotion as the source through the loss $L_{emog}$.
\begin{equation}
\vspace{-0.2em}
\label{eqn:emog}
L_{emog} = \E_{X_s, e_s, h_{sc}}\big[CrossEntropy(C_{e}(G(X_s,h_{sc})), e_s)\big]
\vspace{-0.2em}
\end{equation}
\subsection{Indirect Emotion Supervision}
Incorporation of explicit emotion supervision for the converted samples becomes challenging due to the unavailability of the emotion labels.
Therefore, it becomes pertinent to measure the emotion discrepancy between the source and the converted samples through representations of emotion.
To this end, we propose two ways to measure discrepancies of the emotional content: acoustic features correlated to emotion and deep emotion embeddings. 

 \subsubsection{Emotion-aware Acoustic Feature Loss}
 \label{sec:AF}

We propose acoustic feature loss $L_{af}$, an unsupervised loss computed between the acoustic descriptors of the source and converted samples, as shown in Equation~\ref{eqn:af}, where $AF$ denotes an acoustic feature.
The acoustic features are correlated with emotion and require
\vspace{-0.25em}
 \begin{equation}
L_{af} = \E_{X_s, h_{sc}}\big[\lVert AF(X_s) - AF(G(X_s,h_{sc}))\rVert_1\big]
\label{eqn:af}
\vspace{-0.25em}
 \end{equation}
being differentiable to provide feedback to the network.
Based on~\cite{schuller2009interspeech}, the acoustic descriptors can be categorised into two groups, spectral and non-spectral.
Spectral features add additional information about higher-level harmonics to that already existing in pitch, which provides pertinent cues for the emotional state~\cite{tursunov2019discriminating}.
Many works~\cite{tursunov2019discriminating, busso2012unveiling} report spectral features to be better discriminators in between emotions that have different degree of polarity (valence) but similar intensity (arousal), such as anger and happiness.
The non-spectral features are energy or voicing-related, which are typically prosodic and arousal indicative~\cite{weninger2013acoustics}.
We consider two descriptors from each of the two categories.
All descriptors are extracted over voiced segments using 50\% overlapping windows, to capture the local transients.
\begin{itemize}
\item \textbf{Spectral centroid:} 
Higher spectral centroid values indicate emotions positioned in the upper-right quadrant of the valence-arousal 2D plane, such as \textit{excited} or \textit{happy}~\cite{SchullerWoellmerEybenetal.2009}.
Lower values indicate subdued emotions, such as \textit{sad}.
\item \textbf{Spectral kurtosis:} 
Spectral kurtosis shows the existence of increased energy concentration within specific frequency ranges.
Further, it can detect the series of transients~\cite{antoni2006spectral}, which can make it a good indicator of emotions, especially the ones having subtle intonation changes, such as in the emotion \textit{surprise}.

\item \textbf{Loudness:}
Loudness is an arousal indicative non-spectral feature, which correlates stronger to emotion than root-mean-square energy due to the perceptual A-weighting~\cite{FRUHHOLZ201696}.
Louder sounds elicit stronger emotional responses (high arousal), and vice-versa.
\item \textbf{Change in F0 ($\Delta$F0):}
$\Delta$F0 is a prosodic non-spectral feature, which captures change in intonation, where a considerable change implies stronger emotions, such as \textit{anger} or \textit{excited}~\cite{FRUHHOLZ201696}.
\end{itemize}

\begin{figure}[!t] 
  \centering  \includegraphics[width=0.8\linewidth]{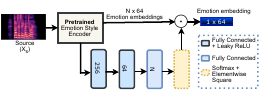}
  \vspace{-2em}
  \caption{Automatic emotion embedding extraction. N denotes the number of emotions classes.}
  \label{fig:emo_selector}
  \vspace{-0.45cm}
\end{figure}
\subsubsection{Emotion Embedding Loss}
Another way of incorporating indirect emotion supervision is through latent emotion representations.
The emotion embedding loss $L_{embed}$ penalises the discrepancy between the latent emotional content of the source and converted samples.
\begin{equation}
\label{eqn:emodeep}
L_{embed} = \E_{X_s, h_{sc}}\big[\lVert Emb(X_s) - Emb(G(X_s,h_{sc}))\rVert_1\big]
\end{equation}
The emotion embedding is obtained by a two-stage training on categorical emotion-labelled data.
At Stage I, the vanilla StarGANv2-VC model is trained for emotion conversion task rather than voice conversion, as shown in Figure~\ref{fig:architecture}.
The emotion style-encoder learns $N \times 64$ embeddings of emotion classes, where $N$ denotes the number of emotion classes.
However, this framework cannot be used in the VC training, as an emotion label (code) is required to generate the emotion embeddings, which is unknown for the converted samples.
Therefore, the pre-trained emotion style-encoder from Stage I is fine-tuned for automatic embedding extraction, as shown in Figure~\ref{fig:emo_selector}.
At Stage II, the pre-trained emotion style-encoder is extended with fully-connected layers and a softmax distribution is generated over all emotions.
Further, the softmax score is element-wise squared to encourage sparsity.
Finally, a dot product between the sparse $1 \times N$ score and the encoder output is performed to produce a $1 \times 64$ dimensional latent emotion representation.
This fine-tuned model is used in the VC training to extract the emotion embeddings from both source and converted samples.
\subsection{Training Objectives}
\label{sec:tobj}
The components in Emo-StarGAN are trained with the proposed emotion-aware losses along with the losses from StarGANv2-VC.
The generator is trained with loss $L_G$ (Equation~\ref{eqn:g}) comprising the proposed emotion classification loss $L_{emod}$, unsupervised emotion-aware losses ($L_{af}$ and $L_{embed}$), and losses from StarGANv2-VC.
\vspace{-0.1em}
\begin{align*}\tag{5}
\label{eqn:g}
& L_G = \underset{G,SE,M}{min} L_{adv} + \lambda_{af}L_{af} + \lambda_{embed}L_{embed} + \\ 
& \qquad \qquad \quad \lambda_{emog}L_{emog} + \lambda_{aspk}L_{aspk} + \lambda_{sty}L_{sty} \\
& \qquad \qquad  \quad- \lambda_{ds}L_{ds} + \lambda_{F0}L_{F0} + \lambda_{asr}L_{asr} + \lambda_{cyc}L_{cyc}
\vspace{-0.1em}
\end{align*}
The losses from StarGANv2-VC: $L_{adv}$ is the typical GAN adversarial loss,
$L_{aspk}$ is the adversarial source speaker classification loss,
$L_{sty}$ ensures that the style representations can be recreated from the generated samples,
$L_{ds}$ is maximised to ensure samples generated from different speaker style-codes sound different,
$L_{F0}$ encourages the generator to produce F0-consistent samples, $L_{asr}$ ensures source and converted samples have the same linguistic content and
$L_{cyc}$ is the cyclic consistency loss, which preserves the non-timbre features of the source.
$\lambda_{af}$, $\lambda_{embed}$, $\lambda_{emog}$, $\lambda_{aspk}$, $\lambda_{sty}$, $\lambda_{ds}$, $\lambda_{F0}$, $\lambda_{asr}$ and $\lambda_{cyc}$ are hyperparameters of the corresponding losses.
The discriminator and classifiers are trained using the objective function shown in Equation~\ref{eqn:d}, where $\lambda_{spk}$ and $\lambda_{emod}$ are hyperparameters for the source speaker classification loss $L_{spk}$ and emotion classification loss $L_{emod}$, respectively.
\begin{align}\tag{6}
\vspace{-0.1em}
\label{eqn:d}
& L_D = \underset{D, C_{e}, C_s}{min} -L_{adv} + \lambda_{emod}L_{emod} + \lambda_{spk}L_{spk}
\vspace{-0.1em}
\end{align}

\begin{table*}[!ht]
\caption{Objective evaluation. Mean and standard deviation (in brackets) reported. Emo-SG denotes Emo-StarGAN. `All Conv.' includes all conversions. Type column denotes special cases, such as source-emotion, source-accent $\rightarrow$ target-accent, source and target same genders (M$\rightarrow$M, F$\rightarrow$F), source and target different genders (M$\rightarrow$F, F$\rightarrow$M) or, `All' including all sub-groups.}
\resizebox{\textwidth}{!}
{
 \begin{tabular}{l|l|ll|ll|ll|ll|ll|ll|ll}
 \toprule
 {\textbf{\specialcell{Source - Target}}} & {\textbf{\specialcell{Type}}} &
 \multicolumn{2}{c|}{\textbf{Acc\textsubscript{orig}} [\%] $\uparrow$}  & \multicolumn{2}{c|}{\textbf{Acc\textsubscript{svm}} [\%] $\uparrow$} & \multicolumn{2}{c|}{\textbf{Embedding MAE} [$\times 10^{2}$]} $\downarrow$ & \multicolumn{2}{c|}{\textbf{PCC} [$\times 10^{2}$] $\uparrow$} & \multicolumn{2}{c|}{\textbf{pMOS} $\uparrow$} & \multicolumn{2}{c|}{\textbf{CER} [\%] $\downarrow$} & \multicolumn{2}{c}{\textbf{EER} [\%] $\uparrow$} \\ 
\cline{3-16}

{} & {} & {\textbf{Baseline}} & {\textbf{Emo-SG}} & {\textbf{Baseline}} & {\textbf{Emo-SG}} & {\textbf{Baseline}} & {\textbf{Emo-SG}} & {\textbf{Baseline}} & {\textbf{Emo-SG}} & {\textbf{Baseline}} & {\textbf{Emo-SG}} & {\textbf{Baseline}} & {\textbf{Emo-SG}} & {\textbf{Baseline}} & {\textbf{Emo-SG}}\\ 
\midrule
{\textbf{All Conv.}} & {All} & {20.2} & {\textbf{72.4}} & {39.4} & {\textbf{87.1}} & {48.9 (11.1)} & {\textbf{40.8} (10.9)} &{78.9 (12.9)} & {\textbf{84.3} (10.6)} & {3.68 (0.41)} & {\textbf{3.72} (0.44)} & {3.42 (8.39)} & {\textbf{2.57} (7.08)} & {49.63} & {\textbf{49.64}}\\
 \midrule
 \multirow{8}{*}{\textbf{\specialcell{ESD \\ \, \, $ \downarrow$\\ESD}}}  & {All} & {19.1} & {\textbf{68.9}} & {20.1} & {\textbf{94.7}} & {43.4 (17.0)} & {\textbf{31.5} (14.4)} &{78.1 (14.0)} & {\textbf{84.9} (10.4)} & {3.75 (0.43)} & {\textbf{3.90} (0.4)} & {4.27 (8.79)} & {\textbf{3.56} (7.75)} & {\textbf{45.86}} & {45.45}\\
 \cmidrule{2-16}
 & {Happy} & {10.7} & {\textbf{69.7}} & {13.1} & {\textbf{85.7}} & {43.3 (17.5)} & {\textbf{31.0} (14.9)} & {76.7 (16.7)} & {\textbf{84.0} (11.7)} & {3.60 (0.45)} & {\textbf{3.76} (0.41)} & {5.21 (9.36)} & {\textbf{4.92} (8.67)}  & {-} & {-} \\
 & {Sad} & {15.7} & {\textbf{96.2}} & {15.7} & {\textbf{96.2}}  & {40.7 (15.3)} & {\textbf{32.4} (14.5)} & {82.3 (12.7)} & {\textbf{87.5} (10.0)} & {3.76 (0.39)} & {\textbf{4.07} (0.38)} & {2.47 (6.15)} & {\textbf{2.38} (6.0)}  & {-} & {-} \\ 
 
 & {Surprise} & {0.0} & {\textbf{16.0}} & {2.1} & {\textbf{97.9}}  & {47.1 (15.5)} & {\textbf{29.5} (11.6)} & {77.6 (11.2)} & {\textbf{85.4} (7.4)} & {3.64 (0.38)} & {\textbf{3.74} (0.35)} & {7.43 (11.17)} & {\textbf{5.16} (9.3)}  & {-} & {-} \\
 
 & {Angry} & {10.6} & {\textbf{94.0}} & {10.6} & {\textbf{95.0}}  & {43.7 (17.8)} & {\textbf{31.1} (14.5)} & {76.1 (14.8)} & {\textbf{85.6} (9.6)} & {3.77 (0.39)} & {\textbf{3.86} (0.34)} & {3.53 (8.02)} & {\textbf{2.96} (7.01)}  & {-} & {-} \\
 
 & {Neutral} & {79.3} & {\textbf{99.3}} & {79.7} & {\textbf{99.3}}  & {40.7 (18.8)} & {\textbf{34.5} (16.5)} & {78.0 (12.8)} & {\textbf{80.4} (12.6)} & {4.08 (0.37)} & {\textbf{4.19} (0.34)} & {\textbf{1.63 }(5.73)} & {1.69 (5.93)}  & {-} & {-} \\
  \cmidrule{2-16}
  & {Different gender} & {18.6} & {\textbf{63.3}} & {19.6} & {\textbf{94.2}} & {50.05 (15.7)} & {\textbf{38.8} (14.0)} & {76.9 (14.7)} & {\textbf{86.7} (9.0)} & {3.71 (0.44)} & {\textbf{3.92} (0.37)} & {4.82 (9.09)}  & {\textbf{2.53} (5.70)} & {-} & {-} \\
& {Same gender} & {19.5} & {\textbf{78.6}} & {21.1} & {\textbf{96.3}} & {37.6 (15.1)} & {\textbf{25.35} (10.3)} & {79.1 (13.2) } & {\textbf{82.7} (11.6)} & {3.78 (0.41)} & {\textbf{3.80} (0.39)} & {3.78 (8.48)} & {\textbf{3.27} (7.02)}  & {-} & {-} \\
 \midrule
 \multirow{8}{*}{\textbf{\specialcell{RAVDESS \\ \, \, \, $\downarrow$ \\ \, ESD}}} & {All} & {27.8} & {\textbf{49.2}} & {41.4} & {\textbf{76.0}} & {52.5 (7.41)} & {\textbf{44.7} (7.66)} & {86.2 (10.8)} & {\textbf{88.0} (9.3)} & {3.44 (0.41)} & {\textbf{3.49} (0.41)} & {4.57 (9.89)} & {\textbf{4.52 }(5.7)} & {50.19} & {\textbf{50.44}} \\

 
 
 

\cmidrule{2-16}
 & {Happy} & {0.0} & {\textbf{14.0}} & {63.0} & {\textbf{96.0}}  & {51.2 (6.5)} & {\textbf{44.2} (5.8)} & {89.0 (7.2)} & {\textbf{90.7} (6.4)} & {3.35 (0.35)} & {\textbf{3.40} (0.33)} & {5.19 (9.42)} & {\textbf{2.95} (6.49)} & {-} & {-} \\
 
 & {Sad} & {0.0} & {\textbf{59.0}} & {9.0} & {\textbf{73.0}}  & {55.1 (9.5)} & {\textbf{44.8} (6.5)} & {76.1 (16.2)} & {\textbf{81.0} (15.0)} & {\textbf{3.20} (0.47)} & {3.13 (0.45)} & {13.11 (12.08)} & {\textbf{8.96} (2.31)} & {-} & {-} \\
 
 & {Surprise} & {\textbf{0.0}} & {\textbf{0.0}} & {4.0} & {\textbf{34.0}}  & {52.3 (6.1)} & {\textbf{41.2} (6.4)} & {\textbf{89.2} (6.2)} & {\textbf{89.2} (7.5)} & {\textbf{3.37} (0.25)} & {3.35 (0.34)} & {4.60 (8.35)} & {\textbf{3.71} (5.45)} & {-} & {-} \\
 
 & {Angry} & {51.0} & {\textbf{93.0}} & {51.0} & {\textbf{93.0}}  & {50.8 (7.5)} & {\textbf{44.6} (9.1)} & {90.7 (5.4)} & {\textbf{91.8} (4.2)} & {3.74 (0.33)} & {\textbf{3.84} (0.31)} & {\textbf{2.00} (8.50)} & {5.65 (3.02)} & {-} & {-} \\
 
 & {Neutral} & {80.0} & {\textbf{88.0}} & {80.0} & {\textbf{88.0}}  & {53.1 (6.2)} & {\textbf{48.8} (8.1)} & {86.1 (8.2)} & {\textbf{87.7} (5.1)} & {3.34 (0.29)} & {\textbf{3.48} (0.37)} & {2.63 (9.07)} & {\textbf{1.28} (3.86)}  & {-} & {-} \\
\cmidrule{2-16}
  & {Different gender} & {33.7} & {\textbf{43.1}} & {46.8} & {\textbf{70.0}} & {54.7 (5.2)} & {\textbf{32.2} (6.3)} & {87.2 (7.1)} & {\textbf{88.6} (6.9)} & {\textbf{3.46} (0.42)} & {\textbf{3.46} (0.40)} & {5.74 (11.26)} & {\textbf{3.75} (2.50)}  & {-} & {-} \\
  & {Same gender} & {24.2} & {\textbf{52.9}} & {38.1} & {\textbf{72.9}} & {51.6 (7.1)} & {\textbf{43.1} (6.1)} & {85.6 (12.5)} & {\textbf{87.6} (10.5)} & {3.46 (0.42)} & {\textbf{3.51} (0.40)} & {\textbf{3.73} (8.86)} & {4.97 (2.35)}  & {-} & {-} \\
\midrule
 \multirow{4}{*}{\textbf{\specialcell{VCTK \\ \, \, \, $\downarrow$ \\ VCTK}}}  & {All} & {-} & {-} & {56.8} & {\textbf{90.6}} & {50.8 (13.0)} & {\textbf{46.4} (12.1)} & {78.4 (11.8)} & {\textbf{83.1} (10.8)}  & {3.51 (0.36)} & {\textbf{3.57} (0.37)}  & {3.27 (8.31)} & {\textbf{1.63} (5.54)}   & {\textbf{50.13}} & {49.90} \\
\cmidrule{2-16}

 
 
 & {British $\rightarrow$ British} & {-} & {-} & {48.9} & {\textbf{91.6}}  & {51.9 (13.4)} & {\textbf{46.9} (12.6)} & {77.5 (10.8)} & {\textbf{82.5} (9.7)}  & {3.53 (0.35)} &\textbf{{3.62}} (0.36)& {4.16 (9.5)} & {\textbf{2.19} (6.45)} & {-} & {-} \\
 
 & {American $\rightarrow$ British} & {-} & {-} & {66.5} & {\textbf{89.9}}  & {50.0 (12.3)} & {\textbf{46.3} (11.7)} & {80.0 (11.6)} & {\textbf{84.4} (10.4)} & {3.55 (0.38)} & {\textbf{3.49} (0.37)} & {2.58 (7.59)} & {\textbf{1.25} (5.07)} & {-} & {-} \\
 
 & {Canadian $\rightarrow$ British} & {-} & {-} & {53.3} & {\textbf{89.9}}  & {50.0 (13.3)} & {\textbf{45.9} (12.0)} & {77.2 (13.8)} & {\textbf{81.5} (13.1)} & {3.52 (0.36)} & {\textbf{3.49} (0.37)} & {2.85 (6.86)} & {\textbf{1.28} (4.24)} & {-} & {-} \\
\cmidrule{2-16}
& {Different gender} & - & - &{55.8} & {\textbf{90.4}} & {48.0 (12.0)} & {\textbf{44.0} (10.0)} &{79.1 (10.8)} &{\textbf{83.5} (9.9)} & {3.45 (0.36)} & {\textbf{3.53} (0.36} & {3.62 (8.53)} & {\textbf{1.76} (5.59)}  & {-} & {-} \\
& {Same gender} & - & - & {59.5} & {\textbf{91.0}} & {51.0 (10.5)} & {\textbf{46.1 }(9.1)}& {77.9 (12.5)} & {\textbf{82.7} (11.4)} & {3.55 (0.36)}& {\textbf{3.61} (0.38)} & {2.99 (8.11)} & {\textbf{1.53} (5.50} & {-} & {-}\\
 \bottomrule
\end{tabular}
}
\label{tab:objective}
\vspace{-0.75em}
\end{table*}
\vspace{-1em}
\section{Experiment and Results}
\subsection{Dataset and Training details}
English utterances from VCTK~\cite{yamagishi2019cstr}, emotional speech dataset (ESD)~\cite{zhou2022emotional} and Ryerson audio-visual database of emotional speech and song (RAVDESS)~\cite{livingstone2018ryerson} datasets are considered.
VCTK has no emotion ground truth, whereas ESD and RAVDESS are labelled with categorical emotions, where we consider five emotion classes common to both, $e$ $\in$ \{happy, sad, anger, neutral, surprise\}.
The utterances are re-sampled to \SI{24}{\kilo\hertz} and randomly split as 0.8/0.1/0.1 (train/validation/test).
All VC models are trained on 10 randomly selected speakers from VCTK and ESD each.

Our model has the same number of trainable model parameters as StarGANv2-VC.
Each model is trained on log mel-spectrograms derived from 2~second audio samples, for 100 epochs with a batch size of 16.
Each training takes around 36 hours on average to complete on A100 (\qty{80}{\giga\byte}).
We use pre-trained F0 and  automatic speech recognition models from~\cite{li2021starganv2vc}.
AdamW optimizer~\cite{loshchilov2018decoupled} is used with a learning rate of $10^{-4}$.
We set $\lambda_{aspk}=0.1, \lambda_{emod}=0.01, \lambda_{emog}=0.01, \lambda_{sty}=1, \lambda_{ds}=1, \lambda_{F0}=5, \lambda_{asr}=1, \lambda_{cyc}=1, \lambda_{embed}=2$ and $\lambda_{af}=2$.
A HiFiGAN~\cite{kong2020hifi} vocoder is trained on the mentioned datasets, which generates one-minute long waveform from the converted mel-spectrogram in 0.1 seconds on the A100.
The emotion conversion model is trained using cross-validation only on ESD, using training split 0.9/0.1 (train/validation), and using the same setup as the VC models.
The best model is selected based on the lowest \acrfull{MAE}.
To evaluate emotion preservation, a Support Vector Machine (SVM) based emotion classifier is trained as in~\cite{seehapoch2013speech} 
on \textit{source} utterances of ESD and RAVDESS.
\vspace{-0.75em}
\subsection{Evaluation Setup}
\vspace{-0.25em}
We evaluate our approach using both objective and subjective measures.
We consider StarGANv2-VC as the \textit{baseline}.
Further, we perform experiments to find the best emotion-preserving acoustic feature $AF_{best}$.
We train our model Emo-StarGAN using the combination of emotion classifier loss, emotion embedding loss and acoustic feature loss using $AF_{best}$.
For all experiments, an equal number of female (F) and male (M) speakers are randomly selected as source and target.
From each of the three datasets, 10 source speakers are considered.
For ESD and RAVDESS, 5 utterances for each source speaker and each emotion in $e$ are selected.
We convert source utterances from ESD using ESD target speakers (ESD$\rightarrow$ESD) for \textit{within-corpus} and RAVDESS$\rightarrow$ESD for \textit{cross-corpus} scenarios.
We select 6 target speakers from ESD, leading to 1500 conversions for each scenario.
For the \textit{inter-accent} conversion use case, VCTK$\rightarrow$VCTK conversion is performed, where 10 utterances from each source speaker and accent group (British, American, and Canadian) and 6 target speakers having British accent are considered, leading to 1800 conversions.

\textbf{Objective Evaluation:}
Emotion preservation is evaluated in four ways: (i) Acc\textsubscript{orig}: SVM classification accuracy, considering the emotion labels of the source utterance provided in the dataset, (ii) Acc\textsubscript{svm}: SVM classification accuracy, considering SVM prediction on source utterance as the emotion ground truth, (iii) Embedding MAE: mean absolute error between the source and converted emotion embedding outputs, (iv) Pitch correlation coefficient (PCC): measures the degree of intonation preservation~\cite{tomashenko2022voiceprivacy}, which provides cues to emotion preservation~\cite{rodero2011intonation}.
The voice quality is measured by predicted mean opinion score (pMOS)~\cite{lo2019mosnet}.
We report the character error rate (CER) using the transcriptions from Whisper \textit{medium-english} model~\cite{radford2022robust}.
Equal error rate (EER) measures anonymisation using the state-of-the-art speaker verification model ECAPA-TDNN~\cite{desplanques2020ecapa}.
For the metrics Acc\textsubscript{orig}, Acc\textsubscript{svm}, PCC, pMOS, and EER higher values indicate better performance, and for Embedding MAE and CER lower values are preferable.

\textbf{Subjective evaluation:}
We consider 100 randomly selected conversions for subjective evaluation as it is expensive and time-consuming to perform for all.
138 online subjects participated in the user study through the Crowdee\footnote{https://www.crowdee.com/} platform.
For emotion preservation assessment, subjects were presented with the source utterance and two options: conversions from baseline and Emo-StarGAN.
Further, they were asked to choose one of the options having similar rhythm, intonation, pauses, stresses and intensity as the source, irrespective of voice quality and the linguistic content.
The subjects were asked to rate on a 5-point scale for naturalness (1: bad to 5: excellent).
For speaker anonymisation, the raters were presented with the converted sample and another utterance from the source speaker, and were asked to rate on a 5-point scale (1: different to 5: similar).
At least three subjects rated each task.
The raters were not informed whether the samples are original or converted. They were further provided with anchoring examples and hidden trapping questions. Raters caught in the latter twice were rejected from evaluations.

\subsection{Results and Discussion}
\noindent\textbf{Selection of Acoustic Feature (AF) and Ablation:}
In order to get $AF_{best}$, we train the baseline with acoustic feature loss, where the $AF$ is replaced with one of the acoustic features mentioned in Section~\ref{sec:AF}.
Among all acoustic features, \textit{spectral kurtosis} preserves emotion the most (30.1\% Acc\textsubscript{orig}, 80.4 PCC), also  outperforming the baseline (19\% Acc\textsubscript{orig}, 78.1 PCC).
The PCC values of the other acoustic features are similar, having range $\numrange{80.0}{80.4}$.
Acc\textsubscript{orig} for the other acoustic features are, spectral centroid (24.1\%), loudness (15.4\%) and $\Delta F0$ (20.1\%), which portrays the spectral features to be more emotion preserving than the non-spectral ones, compliant with~\cite{tursunov2019discriminating}.
The ablation study (Table~\ref{tab:ablation}) shows that the unsupervised loss $L_{embed}$ contributes the most to emotion preservation, even more than the direct supervision by emotion classifier $C_{e}$, this might be attributed to $C_{e}$ suffering from confirmation bias on noisy emotion labels.
Further, we observe that each individual proposed technique preserves emotion more than the baseline.
\begin{table}[h]
\caption{Ablation results. Mean and standard deviation (in brackets) reported. L\textsubscript{af} uses spectral kurtosis as the acoustic feature. Baseline is trained with `none' of the emotion-aware losses.}
\resizebox{0.47\textwidth}{!}
{
 \begin{tabular}{llllll}
 \toprule
     {\textbf{\textbf{Method}}} & {\textbf{Acc\textsubscript{orig}} [\%] $\uparrow$} & 
     {\textbf{PCC [$\times 10^{2}$] $\uparrow$}} & {\textbf{pMOS} $\uparrow$} & {\textbf{CER} [\%] $\downarrow$} & {\textbf{EER} [\%] $\uparrow$}   \\
     \midrule
    {Baseline} &20.2  
    &78.9 (12.9)& 3.68 (0.41)& 3.42 (8.39) & 49.63 \\
    {Emo-StarGAN} &\textbf{72.4}&  
    \textbf{84.3} (10.6)& \textbf{3.72} (0.44)& \textbf{2.57} (7.08) &\textbf{49.64} \\
    \midrule
    {L\textsubscript{embed}}& \textbf{51.0}
    &\textbf{81.3} (12.2) & \textbf{3.90} (0.40) &\textbf{3.12} (7.97)& \textbf{48.09}\\
    {C\textsubscript{e}}& 49.3  
    & 81.0 (12.3)& 3.50 (0.46)& 3.50 (7.76) & 45.83\\
    {L\textsubscript{af}} & 30.1& 
    80.4 (11.8)& 3.89 (0.37)& 5.52 (11.68) & 47.64\\
     \bottomrule
\end{tabular}
}
\label{tab:ablation}
\end{table}

\noindent\textbf{Comparison with Baseline}:
Our method Emo-StarGAN outperforms the baseline with respect to emotion preservation for all scenarios (Table~\ref{tab:objective}), which is also statistically significant (p $<$ 0.001 for paired t-test on PCC and Embedding MAE columns).
The subjective evaluation (Table \ref{tab:subj}) also shows that our model is voted more emotion preserving (72\%) compared to the baseline (28\%). 
\textit{Surprise} is reported as one of the most difficult emotions in speech emotion recognition tasks~\cite{chen2012speech}. 
Our method also achieves lower accuracy for `surprise' compared to other emotions, where Acc\textsubscript{orig} scores for ESD and RAVDESS are only 16\% and 0\% respectively.
However, preservation seems much higher considering Acc\textsubscript{svm} scores, 97.9\% ESD and 34\% for RAVDESS.
Our framework improves emotion preservation significantly for the cross-corpus (RAVDESS$\rightarrow$ESD) scenario with respect to all metrics, especially for \textit{sad}, where the emotion preservation improves from 0\% to 59\% (Acc\textsubscript{orig}), 9\% to 73\% (Acc\textsubscript{svm}), 55.1 to 44.8 (Embedding MAE) and 76\% to 81\% (PCC).
Considering inter-accent cases, our model produces a high Acc\textsubscript{svm} score of 89.9\% for both American~$\rightarrow$~British and Canadian~$\rightarrow$~British conversions, and also improves other quality metrics. 
For both gender conversion cases, similar observations are made.
Our method outperforms the baseline mostly with respect to voice quality, intelligibility, and anonymisation, which is further supported by the subjective results.
The code and demo audio samples can be found online\footnote{https://github.com/suhitaghosh10/emo-stargan.git}.
\begin{table}[h]
\caption{Results of subjective evaluation. Mean and standard deviation (in brackets) reported. Emo-SG denotes Emo-StarGAN. Emotion v. column denotes the number of times a model is preferred over the other. Higher Speaker diss. indicates better anonymisation. }
\vspace{-0.2cm}
\resizebox{0.47\textwidth}{!}
{
 \begin{tabular}{l|ll|ll|ll}
 \toprule
\multirow{2}{*}{\textbf{Type}} & \multicolumn{2}{c|}{\textbf{MOS} $\uparrow$} & \multicolumn{2}{c|}{\textbf{Emotion V.} $\uparrow$} & \multicolumn{2}{c}{\textbf{Speaker Diss.} $\uparrow$} \\ 
\cline{2-7}
{} & {\textbf{Baseline}} & {\textbf{Emo-SG}} & {\textbf{Baseline}} & {\textbf{Emo-SG}} & {\textbf{Baseline}} & {\textbf{Emo-SG}}\\ 
\midrule
{All} &  {4.09 (0.93)} & {\textbf{4.20} (0.93)} & {327} & {\textbf{840}}& {2.4 (1.4)} & {\textbf{2.6} (1.5)} \\
\midrule
{Different gender} &{4.20 (0.94)} & {\textbf{4.24} (0.93)} & {154} & {\textbf{429}}& {2.7 (1.4)} &{\textbf{2.9} (1.5)}\\
{Same gender} &  {4.05 (0.94)} & {\textbf{4.19} (1.01)} & {173} & {\textbf{411}}& {1.9 (1.3)} &{\textbf{2.3} (1.5)} \\
 \bottomrule
\end{tabular}
}
\label{tab:subj}
\end{table}
\vspace{-2em}
\section{Conclusions}
To the best of our knowledge, we propose the first emotion-preserving any-to-many semi-supervised voice conversion framework Emo-StarGAN.
We introduce novel unsupervised acoustic descriptor-based and deep emotion losses, which can be used with any other framework.
Extensive experiments show that Emo-StarGAN preserves emotion significantly better than the state-of-the-art VC method StarGANv2-VC over seen source speakers, cross-corpus conversions, different genders, accents and emotions.
Subjective results show that our method even achieves higher MOS and anonymisation scores.
As future work, we plan to improve the emotion preservation for complex emotions by incorporating losses beneficial to a specific emotion.
Further, we would like to extend the method with emotion embeddings learned from multi-label and arousal-valence labelled datasets.
\section{Acknowledgements}
This research has been partly funded by the Federal Ministry of Education and Research of Germany in the project Emonymous (project number S21060A) and partly funded by the Volkswagen Foundation in the project AnonymPrevent (AI-based Improvement of Anonymity for Remote Assessment, Treatment and Prevention against Child Sexual Abuse).

\bibliographystyle{IEEEtran}
\bibliography{thegreatstargan}

\begin{thebibliography}{10}
\providecommand{\url}[1]{#1}
\csname url@samestyle\endcsname
\providecommand{\newblock}{\relax}
\providecommand{\bibinfo}[2]{#2}
\providecommand{\BIBentrySTDinterwordspacing}{\spaceskip=0pt\relax}
\providecommand{\BIBentryALTinterwordstretchfactor}{4}
\providecommand{\BIBentryALTinterwordspacing}{\spaceskip=\fontdimen2\font plus
\BIBentryALTinterwordstretchfactor\fontdimen3\font minus \fontdimen4\font\relax}
\providecommand{\BIBforeignlanguage}[2]{{%
\expandafter\ifx\csname l@#1\endcsname\relax
\typeout{** WARNING: IEEEtran.bst: No hyphenation pattern has been}%
\typeout{** loaded for the language `#1'. Using the pattern for}%
\typeout{** the default language instead.}%
\else
\language=\csname l@#1\endcsname
\fi
#2}}
\providecommand{\BIBdecl}{\relax}
\BIBdecl

\bibitem{wienrich2021trustworthiness}
C.~Wienrich, C.~Reitelbach, and A.~Carolus, ``The trustworthiness of voice assistants in the context of healthcare investigating the effect of perceived expertise on the trustworthiness of voice assistants, providers, data receivers, and automatic speech recognition,'' \emph{Frontiers in Computer Science}, vol.~3, p. 685250, 2021.

\bibitem{HaaseHCII2022}
M.~Haase, J.~Krüger, and I.~Siegert, ``User perspective on anonymity in voice assistants,'' in \emph{Proc. of the HCI International}, 2023, p. s.p.

\bibitem{RUB35}
D.~Kumar, R.~Paccagnella, P.~Murley, E.~Hennenfent, J.~Mason, A.~Bates, and M.~Bailey, ``Skill squatting attacks on {Amazon Alexa},'' in \emph{27th {USENIX} Security Symposium}, Baltimore, USA, 2018, pp. 33--47.

\bibitem{RUB66}
N.~{Zhang}, X.~{Mi}, X.~{Feng}, X.~{Wang}, Y.~{Tian}, and F.~{Qian}, ``Dangerous skills: Understanding and mitigating security risks of voice-controlled third-party functions on virtual personal assistant systems,'' in \emph{IEEE Symposium on Security and Privacy}, 2019, pp. 1381--1396.

\bibitem{Kobayashi2016TheNV}
K.~Kobayashi, S.~Takamichi, S.~Nakamura, and T.~Toda, ``{The NU-NAIST Voice Conversion System for the Voice Conversion Challenge 2016},'' in \emph{Proc. Interspeech 2016}, 2016, pp. 1667--1671.

\bibitem{helander2011voice}
E.~Helander, H.~Sil{\'e}n, T.~Virtanen, and M.~Gabbouj, ``Voice conversion using dynamic kernel partial least squares regression,'' \emph{IEEE transactions on audio, speech, and language processing}, vol.~20, no.~3, pp. 806--817, 2011.

\bibitem{takashima2013exemplar}
R.~Takashima, T.~Takiguchi, and Y.~Ariki, ``Exemplar-based voice conversion using sparse representation in noisy environments,'' \emph{IEICE Transactions on Fundamentals of Electronics, Communications and Computer Sciences}, vol.~96, no.~10, pp. 1946--1953, 2013.

\bibitem{jin2016cute}
Z.~Jin, A.~Finkelstein, S.~DiVerdi, J.~Lu, and G.~J. Mysore, ``Cute: A concatenative method for voice conversion using exemplar-based unit selection,'' in \emph{Proc. of the IEEE ICASSP}.\hskip 1em plus 0.5em minus 0.4em\relax IEEE, 2016, pp. 5660--5664.

\bibitem{liu2020voice}
R.~Liu, X.~Chen, and X.~Wen, ``Voice conversion with transformer network,'' in \emph{Proc. of the IEEE ICASSP}, 2020, pp. 7759--7759.

\bibitem{lian2020arvc}
Z.~Lian, Z.~Wen, X.~Zhou, S.~Pu, S.~Zhang, and J.~Tao, ``{ARVC}: An auto-regressive voice conversion system without parallel training data.'' in \emph{INTERSPEECH}, 2020, pp. 4706--4710.

\bibitem{sun2016phonetic}
L.~Sun, K.~Li, H.~Wang, S.~Kang, and H.~Meng, ``Phonetic posteriorgrams for many-to-one voice conversion without parallel data training,'' in \emph{2016 IEEE International Conference on Multimedia and Expo (ICME)}.\hskip 1em plus 0.5em minus 0.4em\relax IEEE, 2016, pp. 1--6.

\bibitem{liu2021any}
S.~Liu, Y.~Cao, D.~Wang, X.~Wu, X.~Liu, and H.~Meng, ``Any-to-many voice conversion with location-relative sequence-to-sequence modeling,'' \emph{IEEE/ACM Transactions on Audio, Speech, and Language Processing}, vol.~29, pp. 1717--1728, 2021.

\bibitem{cao2014prosodic}
H.~Cao, {\v{S}}.~Be{\v{n}}u{\v{s}}, R.~C. Gur, R.~Verma, and A.~Nenkova, ``Prosodic cues for emotion: analysis with discrete characterization of intonation,'' \emph{Speech prosody (Urbana, Ill.)}, vol. 2014, p. 130, 2014.

\bibitem{wu2020vqvc+}
D.-Y. Wu, Y.-H. Chen, and H.-y. Lee, ``{VQVC}+: One-shot voice conversion by vector quantization and {U-Net} architecture,'' \emph{Proc. Interspeech 2020}, pp. 4691--4695, 2020.

\bibitem{tang2022avqvc}
H.~Tang, X.~Zhang, J.~Wang, N.~Cheng, and J.~Xiao, ``Avqvc: One-shot voice conversion by vector quantization with applying contrastive learning,'' in \emph{Proc. of the IEEE ICASSP}, 2022, pp. 4613--4617.

\bibitem{sisman2020overview}
B.~Sisman, J.~Yamagishi, S.~King, and H.~Li, ``An overview of voice conversion and its challenges: From statistical modeling to deep learning,'' \emph{IEEE/ACM Transactions on Audio, Speech, and Language Processing}, vol.~29, pp. 132--157, 2020.

\bibitem{sakamoto2021stargan}
S.~Sakamoto, A.~Taniguchi, T.~Taniguchi, and H.~Kameoka, ``{StarGAN-VC+ASR: StarGAN-Based Non-Parallel Voice Conversion Regularized by Automatic Speech Recognition},'' in \emph{Proc. Interspeech 2021}, 2021, pp. 1359--1363.

\bibitem{kaneko2018cyclegan}
T.~Kaneko and H.~Kameoka, ``{CycleGAN-VC}: Non-parallel voice conversion using cycle-consistent adversarial networks,'' in \emph{2018 26th IEEE EUSIPCO}, 2018, pp. 2100--2104.

\bibitem{li2021starganv2vc}
Y.~A. Li, A.~Zare, and N.~Mesgarani, ``{StarGANv2-VC: A Diverse, Unsupervised, Non-Parallel Framework for Natural-Sounding Voice Conversion},'' in \emph{Proc. Interspeech 2021}, 2021, pp. 1349--1353.

\bibitem{kum2019joint}
S.~Kum and J.~Nam, ``Joint detection and classification of singing voice melody using convolutional recurrent neural networks,'' \emph{Applied Sciences}, vol.~9, no.~7, p. 1324, 2019.

\bibitem{walczyna2023overview}
T.~Walczyna and Z.~Piotrowski, ``Overview of voice conversion methods based on deep learning,'' \emph{Applied Sciences}, vol.~13, no.~5, p. 3100, 2023.

\bibitem{schuller2009interspeech}
B.~Schuller, S.~Steidl, and A.~Batliner, ``{The INTERSPEECH 2009 emotion challenge},'' in \emph{Proc. Interspeech 2009}, 2009, pp. 312--315.

\bibitem{tursunov2019discriminating}
A.~Tursunov, S.~Kwon, and H.-S. Pang, ``Discriminating emotions in the valence dimension from speech using timbre features,'' \emph{Applied Sciences}, vol.~9, no.~12, p. 2470, 2019.

\bibitem{busso2012unveiling}
C.~Busso and T.~Rahman, ``Unveiling the acoustic properties that describe the valence dimension,'' in \emph{Thirteenth Annual Conference of the International Speech Communication Association}, September 2012, pp. 1179--1182.

\bibitem{weninger2013acoustics}
F.~Weninger, F.~Eyben, B.~W. Schuller, M.~Mortillaro, and K.~R. Scherer, ``On the acoustics of emotion in audio: what speech, music, and sound have in common,'' \emph{Frontiers in psychology}, vol.~4, p. 292, 2013.

\bibitem{SchullerWoellmerEybenetal.2009}
B.~Schuller, M.~W{\"o}llmer, F.~Eyben, and G.~Rigoll, ``Prosodic, spectral or voice quality? feature type relevance for the discrimination of emotion pairs,'' in \emph{The role of prosody in affective speech}, S.~Hancil, Ed., 2009, pp. 285--307.

\bibitem{antoni2006spectral}
J.~Antoni, ``The spectral kurtosis: a useful tool for characterising non-stationary signals,'' \emph{Mechanical systems and signal processing}, vol.~20, no.~2, pp. 282--307, 2006.

\bibitem{FRUHHOLZ201696}
S.~Frühholz, W.~Trost, and S.~A. Kotz, ``The sound of emotions—towards a unifying neural network perspective of affective sound processing,'' \emph{Neuroscience \& Biobehavioral Reviews}, vol.~68, pp. 96--110, 2016.

\bibitem{yamagishi2019cstr}
J.~Yamagishi, C.~Veaux, K.~MacDonald \emph{et~al.}, ``{CSTR VCTK Corpus}: English multi-speaker corpus for {CSTR} voice cloning toolkit (version 0.92),'' \emph{University of Edinburgh. The Centre for Speech Technology Research (CSTR)}, 2019.

\bibitem{zhou2022emotional}
K.~Zhou, B.~Sisman, R.~Liu, and H.~Li, ``Emotional voice conversion: Theory, databases and {ESD},'' \emph{Speech Communication}, vol. 137, pp. 1--18, 2022.

\bibitem{livingstone2018ryerson}
S.~R. Livingstone and F.~A. Russo, ``The {Ryerson} audio-visual database of emotional speech and song (ravdess): A dynamic, multimodal set of facial and vocal expressions in north american english,'' \emph{PloS one}, vol.~13, no.~5, p. e0196391, 2018.

\bibitem{loshchilov2018decoupled}
I.~Loshchilov and F.~Hutter, ``Decoupled weight decay regularization,'' in \emph{International Conference on Learning Representations}, 2019.

\bibitem{kong2020hifi}
J.~Kong, J.~Kim, and J.~Bae, ``{H}i{F}i-{GAN}: Generative adversarial networks for efficient and high fidelity speech synthesis,'' \emph{Advances in Neural Information Processing Systems}, vol.~33, pp. 17\,022--17\,033, 2020.

\bibitem{seehapoch2013speech}
T.~Seehapoch and S.~Wongthanavasu, ``Speech emotion recognition using support vector machines,'' in \emph{2013 5th international conference on Knowledge and smart technology (KST)}.\hskip 1em plus 0.5em minus 0.4em\relax IEEE, 2013, pp. 86--91.

\bibitem{tomashenko2022voiceprivacy}
\BIBentryALTinterwordspacing
N.~Tomashenko, X.~Wang, X.~Miao, H.~Nourtel, P.~Champion, M.~Todisco, E.~Vincent, N.~Evans, J.~Yamagishi, and J.~F. Bonastre. {The VoicePrivacy 2022 Challenge Evaluation Plan}. Visited on 2023-03-03. [Online]. Available: \url{https://arxiv.org/pdf/2203.12468.pdf}
\BIBentrySTDinterwordspacing

\bibitem{rodero2011intonation}
E.~Rodero, ``Intonation and emotion: influence of pitch levels and contour type on creating emotions,'' \emph{Journal of voice}, vol.~25, no.~1, pp. e25--e34, 2011.

\bibitem{lo2019mosnet}
C.-C. Lo, S.-W. Fu, W.-C. Huang, X.~Wang, J.~Yamagishi, Y.~Tsao, and H.-M. Wang, ``{MOSNet}: Deep learning-based objective assessment for voice conversion,'' \emph{Proc. Interspeech 2019}, pp. 1541--1545, 2019.

\bibitem{radford2022robust}
A.~Radford, J.~W. Kim, T.~Xu, G.~Brockman, C.~McLeavey, and I.~Sutskever, ``Robust speech recognition via large-scale weak supervision,'' OpenAI: San Francisco, CA, USA, Tech. Rep., 2022.

\bibitem{desplanques2020ecapa}
B.~Desplanques, J.~Thienpondt, and K.~Demuynck, ``{ECAPA-TDNN}: {E}mphasized channel attention, propagation and aggregation in tdnn based speaker verification,'' \emph{Proc. Interspeech 2020}, pp. 3830--3834, 2020.

\bibitem{chen2012speech}
L.~Chen, X.~Mao, Y.~Xue, and L.~L. Cheng, ``Speech emotion recognition: Features and classification models,'' \emph{Digital signal processing}, vol.~22, no.~6, pp. 1154--1160, 2012.

\end{thebibliography}

\end{document}